# On the theory of Double Quantum NMR in polymer systems: the second cumulant approximation for many spin I=1/2 systems

N. Fatkullin,[1] C. Mattea,[2] S. Stapf[2]

[1]Institute of Physics, Kazan Federal University, Kazan, 420008, Tatarstan, Russia, [2]Technische Universität Ilmenau, Dept. Technical Physics II, 98684 Ilmenau, Germany.

General analytical expressions for Double Quantum Nuclear Magnetic Resonance (NMR) kinetic curves of many-spin I=1/2 systems are derived with an accuracy of the second cumulant approximation. The expressions obtained exactly describe the initial part of the kinetic curves and provide a reasonable approximation up to times of about twice the effective spin-relaxation time. For the case when the system contains two isolated spins, this result exactly reproduces known expressions. In the case of polymer melts, the intermolecular magnetic dipole-dipole interactions significantly influence the time dependence of the DQ NMR kinetic curves.

1. Introduction.

Methods of double-quantum Nuclear Magnetic Resonance (NMR) spectroscopy nowadays assume an essential place in polymer physics as a valuable tool for investigating the structure and dynamics of various polymer systems such as polymer melts in bulk, polymer melts confined in porous media, polymer networks (see, for example, [1-11] and references therein). The typical time scale at which this NMR method is operating in discussed polymer systems is several milliseconds. The theory of the method on which structural and dynamical information is obtained from experimentally observed kinetic curves is essentially based on the so-called -spin – pair interaction approximation. This is equivalent to the assumption that – on experimentally relevant time scales – the specific spin kinetics reflected by DQ NMR is governed mainly by magnetic dipole-dipole interactions between two nearest spins. However, most recent developments in the theory and experiments of proton NMR in polymer melts [14-20] show that at frequencies below several MHz, corresponding to times longer than $10^{-7} s$, contributions from intermolecular magnetic dipole-dipole interactions are not negligible any longer, and actually opens up new opportunities for development of proton NMR in polymer melts. The importance of the intermolecular magnetic dipole-dipole interactions was recently confirmed by computer simulations [21]. Most recent experiments of DQ NMR in polymer melts also appear to indicate a strong effect of intermolecular magnetic dipole-dipole interactions on the observed kinetic curves [22].

For this reason we consider it important to extend the existing theory of DQ NMR in polymer melts with an aim to take quantitatively into account effects of the magnetic dipole-dipole interactions onto the DQ NMR kinetic curves. The main goal of this paper is to derive analytical expressions for the DQ signal evolution which take into account intermolecular magnetic dipole-dipole interactions, thus providing a possibility for estimating the relative contributions to the time dependence of the DQ signal.

## 2. Theoretical part.

The Double Quantum proton NMR can, in most general terms, be characterized as a response of the spin system to the particular DQ pulse sequence (see details in [8,13,23]) which effectively transforms the Hamiltonian of the magnetic dipole-dipole interactions in the rotating frame to the so-called DQ Hamiltonian. A considerable number of different pulse sequences have been employed in the literature, which have in common to create a DQ Hamiltonian, and return it into observable single-quantum magnetization. In the following, we assume that the DQ Hamiltonian is created by the so-called static Baum-Pines sequence (see details, for example in [8,23]). This is actually not a serious restriction since alternative pulse sequences generally create essentially the same effective Hamiltonian, save for possibly different numerical coefficients.

The time intervals during which the DQ pulse sequence is acting on the spin system consists of two equivalent parts of equal duration. The first part is named "excitation" and its duration is counted from time moment zero till the experimentally controlled time $\tau_{DQ}$. The second part is named "reconversion" and its duration is from time moment $\tau_{DQ}$ till the time moment $2\tau_{DQ}$.

The effective DQ magnetic dipole-dipole interaction Hamiltonian which is responsible for the spin kinetics during the "excitation" part of the experiment reads:

$$\hat{H}_{DQ} = \sum_{i<j} \hbar \omega_{eff}^{ij}(t)\left(\hat{I}_i^x \hat{I}_j^x - \hat{I}_i^y \hat{I}_j^y\right). \tag{1}$$

The parameter $\omega_{eff}^{ij}(t)$ describes an effective strength of dipole-dipole coupling protons with numbers $i$ and $j$ during the time interval when radiofrequency pulses are acting and is given by the following expression:

$$\omega_{eff}^{ij}(t) = \frac{1}{2}\frac{\gamma_H^2 \hbar \left(3\cos^2\left(\theta_{ij}(t)\right)-1\right)}{r_{ij}^3(t)}, \tag{2}$$

where $r_{ij}(t)$ is the distance between interacting spins at the time moment $t$ and $\theta_{ij}(t)$ is the angle between direction $Z$ defined as the direction along which the external magnetic field is aligned, and the vector connecting discussed spins. With the transition to the rotating frame, also a transition to the Dirac representation is assumed, which gives the time dependence to $r_{ij}(t)$ and $\theta_{ij}(t)$, over the phase variables connected with the lattice variables.

Note that the effective Hamiltonian (1), in variance with the conventional magnetic dipole-dipole Hamiltonian, can induce only coherent two-spin transitions, i.e. it generates a simultaneous "up" or "down" projection of interacting spins relative to the $Z$ axis.

The propagator which governs the time evolution of the density matrix of the total system containing spins and lattice is determined by the following unitary operator, which in terms of the Dyson chronological exponent can be written as (see, for example, [24]):

$$U^{ex}(\tau_{DQ}) = \hat{T}\exp\left(-i\int_0^{\tau_{DQ}} \sum_{i<j} \omega_{eff}^{ij}(t_1)\left(\hat{I}_i^x\hat{I}_j^x - \hat{I}_i^y\hat{I}_j^y\right)dt_1\right). \tag{3}$$

During the second reconversion period the spin system in the rotating frame is evolved either by the effective Hamiltonian (1), when phase shifts of radiofrequency pulses between excitation and reconversion periods equal to $\Delta\varphi = 0, 180°, 360°$, or by its inverse, i.e. multiplied by -1, if those phase shifts are equal to $\Delta\varphi = 90°, 270°$. The propagator for this period of evolution can be written as:

$$U_n^{rec}(\tau_{DQ}) = \hat{T}\exp\left((-1)^{n+1} i \int_{\tau_{DQ}}^{2\tau_{DQ}} \sum_{i<j} \omega_{ef}^{ij}(t_1)\left(\hat{I}_i^x\hat{I}_j^x - \hat{I}_i^y\hat{I}_j^y\right)dt_1\right), \tag{4}$$

where $n = 0, 2, \ldots$ for $\Delta\varphi = 0, 180°, 360°$ and $n = 1, 3, \ldots$ for $\Delta\varphi = 90°, 270°$.

The total propagator which governs the evolution of the density matrix of the total system during excitation and evolution periods is the product of the propagators (3) and (4):

$$U_n^{DQ} = U_n^{rec}(\tau_{DQ})U^{ex}(\tau_{DQ}). \tag{5}$$

During a DQ experiment the spin system is rotated by an angle $\pi/2$ about the X axis in the rotating frame by the additional radio frequency pulse and the Y component of the total spin is measured as the function of $\tau_{DQ}$. This quantity can be calculated, in accordance with standard ways of statistical mechanics, as follows:

$$A_n(2\tau_{DQ}) = -\frac{\beta\hbar\omega_0}{(2I+1)^{N_s}} \left\langle Tr\left(\hat{I}^y \hat{P}^x_{\pi/2} U^{DQ}_n \hat{I}^z U^{*DQ}_n \hat{P}^x_{-\pi/2}\right)\right\rangle, \quad (6)$$

where the bracket $\langle ... \rangle$ represents the average over equilibrium distribution of the lattice variables which in the present case are the phase variables of the polymer nucleus. The high temperature approximation for the equilibrium spin density matrix is assumed, $\hat{I}^\alpha = \sum_m \hat{I}^\alpha_m$ is the $\alpha$-th component of the total spin, $\hat{P}^x_{\pi/2} = \exp\left(-i\frac{\pi}{2}\hat{I}^x\right)$ is the operator describing the action of the radio frequency pulse, $N_s$ is the total number of spins in the system with the resonance frequency $\omega_0$, $\hbar$ is Planck's constant divided by $2\pi$, $\beta$ is the inverse temperature, and the trace operation $Tr(...)$ is performed over the spin variables, $U^{*DQ}_n = \left(U^{DQ}_n\right)^{-1}$.

Inside the trace operation one can commute two operators, therefore the action of the radio frequency operators $\hat{P}^x_{\pi/2}$ and $\hat{P}^x_{-\pi/2}$ can be transferred to the operator $\hat{I}^y$. Then the expression (6) can be rewritten in the following way:

$$A_n(2\tau_{DQ}) = \frac{\beta\hbar\omega_0}{(2I+1)^{N_s}} \left\langle Tr\left(\hat{I}^z U^{DQ}_n \hat{I}^z U^{*DQ}_n\right)\right\rangle. \quad (7)$$

The expression (7) is exact with an accuracy given by which the effective DQ Hamiltonian is created by the radiofrequency pulses during the excitation and reconversion periods. For the following evolution, however, one should be able to treat the Dyson chronological exponents, which, in the general case, remains an unresolved problem of statistical mechanics. For the limit of the classical rigid lattice, however, frequencies $\omega_{eff}^{ij}(t)$ do not depend on time and the Dyson chronological exponents in expressions (3) and (4) become equal to the usual exponents. Moreover, the propagators $U^{ex}(\tau_{DQ})$ and $U_n^{rec}(\tau_{DQ})$ commute with each other and the total propagator can be written as:

$$U_n^{DQ} = U_n^{rec}(\tau_{DQ}) U^{ex}(\tau_{DQ}) = \exp\left(-i \sum_{i<j} \tilde{\varphi}_{ij}^n \left( \hat{I}_i^x \hat{I}_j^x - \hat{I}_i^y \hat{I}_j^y \right) \right), \qquad (8)$$

where

$$\tilde{\varphi}_{ij}^n = \varphi_{ij}^{ex} + (-1)^n \varphi_{ij}^{rec}$$
$$\varphi_{ij}^{ex} = \int_0^{\tau_{DQ}} \omega_{eff}^{ij}(t_1) dt_1 \quad , \quad \varphi_{ij}^{rec} = \int_{\tau_{DQ}}^{2\tau_{DQ}} \omega_{eff}^{ij}(t_1) dt_1. \qquad (9)$$

In the general case, of course, relation (8) is an approximation. Using direct expansion of expressions (5) and (8) into a Taylor series, and following standard but tedious calculations with inserting them into expression (7), one can see that the approximation (8) reproduces correctly the first two moments, i.e. the contributions of the second order or, respectively, $\varphi_{ij}^{rec,ex}$. The propagator (8) is still too complex for an analytical treatment; we will therefore use another approximation which also reproducescorrectly the second moment for the expression (7), i.e. is an equivalent to the second cumulant approximation or, what is entirely equivalent, to the Anderson-Weiss approximation:

$$U_n^{DQ} \simeq \exp\left(-i\sum_{i<j} \tilde{\varphi}_{ij}^n \left(\hat{I}_i^x \hat{I}_j^x - \hat{I}_i^y \hat{I}_j^y\right)\right) \simeq$$
$$\simeq \exp\left(-i\sum_{i<j} \tilde{\varphi}_{ij}^n \hat{I}_i^x \hat{I}_j^x\right) \exp\left(i\sum_{i<j} \tilde{\varphi}_{ij}^n \hat{I}_i^y \hat{I}_j^y\right) \quad (10)$$

Then substituting the approximation (10) into (7) and exploiting commutativity of operators inside the trace operation, (7) can be rewritten as

$$A_n(2\tau_{DQ}) = \frac{\beta\hbar\omega_0}{(2I+1)^{N_s}} \left\langle Tr\left(\left(U_n^{*xx}(2\tau_{DQ})\hat{I}^z U_n^{xx}(2\tau_{DQ})\right)\left(U_n^{*yy}(2\tau_{DQ})\hat{I}^z U_n^{yy}(2)\tau_{DQ}\right)\right)\right\rangle, \quad (11)$$

where

$$U_n^{*xx}(2\tau_{DQ})\hat{I}^z U_n^{xx}(2\tau_{DQ}) = \exp\left(i\sum_{i<j} \tilde{\varphi}_{ij}^n \hat{I}_i^x \hat{I}_j^x\right) \sum_k \hat{I}_k^z \exp\left(-i\sum_{i<j} \tilde{\varphi}_{ij}^n \hat{I}_i^x \hat{I}_j^x\right)$$
$$U_n^{*yy}(2\tau_{DQ})\tau_{DQ}\hat{I}^z U_n^{yy}(2\tau_{DQ}) = \exp\left(i\sum_{i<j} \tilde{\varphi}_{ij}^n \hat{I}_i^y \hat{I}_j^y\right) \sum_k \hat{I}_k^z \exp\left(-i\sum_{i<j} \tilde{\varphi}_{ij}^n \hat{I}_i^y \hat{I}_j^y\right) \quad (12)$$

The factors on the right-hand side of the expressions (12) can be calculated exactly:

$$U_n^{*xx}(2\tau_{DQ})\hat{I}^z U_n^{xx}(2\tau_{DQ}) = \sum_k \left(\hat{I}_k^z \cos\left(\sum_i \tilde{\varphi}_{ik}^n \hat{I}_i^x\right) - \hat{I}_k^y \sin\left(\sum_i \tilde{\varphi}_{ik}^n \hat{I}_i^x\right)\right)$$
$$U_n^{*yy}(2\tau_{DQ})\hat{I}^z U_n^{yy}(2\tau_{DQ}) = \sum_k \left(\hat{I}_k^z \cos\left(\sum_i \tilde{\varphi}_{ik}^n \hat{I}_i^y\right) - \hat{I}_k^x \sin\left(\sum_i \tilde{\varphi}_{ik}^n \hat{I}_i^y\right)\right) \quad (13)$$

Substituting expressions (13) into (11) one obtains the following:

$$A_n(2\tau_{DQ}) = \frac{\beta\hbar\omega_0}{(2I+1)^{N_s}} \left\{ \sum_{k,m} \left[ \begin{array}{l} Tr\left(\hat{I}_k^z \cos\left(\sum_i \tilde{\varphi}_{ik}^n \hat{I}_i^x\right) \hat{I}_m^z \cos\left(\sum_j \tilde{\varphi}_{jm}^n \hat{I}_j^y\right)\right) + \\ + Tr\left(\hat{I}_k^y \sin\left(\sum_i \tilde{\varphi}_{ik}^n \hat{I}_i^x\right) \hat{I}_m^x \sin\left(\sum_j \tilde{\varphi}_{jm}^n \hat{I}_j^y\right)\right) \end{array} \right] \right\} \quad (14)$$

When dealing with proton resonance, i.e. with spins ½ as will be done throughout this paper, the following additional exact relations hold:

$$\exp(ia\hat{I}_i^\alpha) = \cos(a/2) + 2i\hat{I}_i^\alpha \sin(a/2) \tag{13}$$

Then using Euler formula for trigonometric functions, (13) can be evaluated to the following:

$$A_n(2\tau_{DQ}) = \frac{\beta\hbar\omega_0}{4}\left\langle\sum_k\prod_i\cos^2(\tilde{\varphi}_{ik}^n/2)\right\rangle - \\ -\frac{\beta\hbar\omega}{4}\sum_{k,m}\left\langle\sin^2(\tilde{\varphi}_{km}^n/2)\prod_i\cos(\tilde{\varphi}_{ik}^n/2)\cos(\tilde{\varphi}_{im}^n/2)\right\rangle. \tag{14}$$

For the system of spin pairs, when $i,k,m=1,2$, our expression (14) exactly recovers the known result:

$$A_n^{pair}(2\tau_{DQ}) = \frac{\beta\hbar\omega_0}{2}\left\langle\cos(\tilde{\varphi}_{12}^n)\right\rangle. \tag{15}$$

As was already mentioned, (14) reproduces the first two moments of the exact expression (7). Without losing this accuracy, (14) can be rewritten as:

$$A_n(2\tau_{DQ}) = \frac{\beta\hbar\omega_0}{4}\sum_k\left\langle\prod_i\cos(\tilde{\varphi}_{ik}^n)\right\rangle. \tag{16}$$

This can be demonstrated by inserting the following identity

$$\cos(\tilde{\varphi}_{ik}^n) = \cos^2(\tilde{\varphi}_{ik}^n/2) - \sin^2(\tilde{\varphi}_{ik}^n/2) \tag{17}$$

into (14), performing multiplication and keeping terms only up to first order of $\sin^2(\tilde{\varphi}_{ik}^n/2)$.

The normalized DQ intensity is, by definition, given by the following combination:

$$I_{nDQ}(\tau_{DQ}) = \frac{1}{2}\frac{A_1(2\tau_{DQ}) - A_0(2\tau_{DQ})}{A_1(2\tau_{DQ})}. \qquad (18)$$

Making the approximation (16) the normalized DQ then intensity reads

$$I_{nDQ}(\tau_{DQ}) = \frac{1}{2}\frac{\sum_k \left(\left\langle \prod_i \cos(\varphi_{ik}^{ex} - \varphi_{ik}^{rec})\right\rangle - \left\langle \prod_i \cos(\varphi_{ik}^{ex} + \varphi_{ik}^{rec})\right\rangle\right)}{\sum_k \left\langle \prod_i \cos(\varphi_{ik}^{ex} - \varphi_{ik}^{rec})\right\rangle}. \qquad (19)$$

For small phase angles we can approximate:

$$\cos(\varphi) \approx 1 - \frac{1}{2}\varphi^2 \approx \exp\left\{-\frac{1}{2}\varphi^2\right\}. \qquad (20)$$

The expression (16) can thus be approximated as:

$$A_n(2\tau_{DQ}) = \frac{\beta\hbar\omega_0}{4}\sum_k \exp\left\{-\frac{1}{2}\sum_i \langle \tilde{\varphi}_{ik}^{n\,2}\rangle\right\}. \qquad (21)$$

The normalized DQ intensity then can be obtained from expressions (19) and (21) now reads:

$$I_{nDQ}(\tau_{DQ}) = \frac{\sum_k \sinh\left(\sum_i \langle \varphi_{ik}^{ex}\varphi_{ik}^{rec}\rangle\right)}{\sum_k \left(\exp\left\{\sum_i \langle \varphi_{ik}^{ex}\varphi_{ik}^{rec}\rangle\right\}\right)}. \qquad (22)$$

If all spins are equivalent and the quantities inside the sums in the right-hand part of (22) do not depend on the index k, it can simplified as follows:

$$I_{nDQ}(\tau_{DQ}) = \frac{1}{2}\left(1 - \exp\left\{-\frac{2}{N_s}\sum_{i,k} \langle \varphi_{ik}^{ex}\varphi_{ik}^{rec}\rangle\right\}\right). \qquad (23)$$

## 3. Discussion.

Expressions (14), (16), (19), (22) and (23) represent the main theoretical results of this paper. They resolve, in a very general form, the problem of calculating DQ kinetic curves for DQ NMR for many-spin systems, when the effective DQ Hamiltonian is described by the expression (1). For the derivation of mentioned expressions we have used two actually equivalent approximations:

The first one is made by using the expression (8). It is exact for a rigid classical lattice, i.e. for the case when the thermal motions of spin-bearing nuclei are negligibly smaller than the average distances between them. For the classical liquid state, expression (8) is an approximation which exactly reproduces the first two moments of the exact expression (7). This statement can be checked by direct, bulky calculations.

The second approximation (10) again reproduces exactly the first two moments. Note that actually both discussed approximations like the Anderson-Weiss approach reproduce the initial time dependence up to $\tau_{DQ} \ll T_2^{eff}$, where $T_2^{eff}$ is the effective spin-spin relaxation time, i.e. the time during which the free induction signal – which is generally of non-exponential nature – decays to 1/e of its initial value. Expressions (14) and (16) are equivalent for short times $\tau_{DQ} \ll T_2^{eff}$, although one of them can be more successful in describing real experimental data for longer times. It is, however, not possible to predict, solely on theoretical arguments, which of the expressions (18) or (20) is more accurate for longer times. Because the expression (16) is more compact, we will focus our attention to it in the remainder of this paper. Both discussed expressions reproduce exactly the well-known results for DQ kinetic curves for pair spins given by (15). Actually it is this part of the DQ NMR kinetic curves which is most important for the application of this method to the investigation of polymer dynamics in melts. For times $t \gg T_2^{eff}$ the exact expression (7) would be mainly determined for the four-, six- (and higher, even-numbered) body particles dynamical correlations, which on the present level of development of the liquid state in general, and polymer melts in particular, are intractable.

Note also that the expressions obtained above actually would give reasonable results even at times $t \leq 2T_2^{eff}$. The main effect which leads to deviations by using approximations (8) and (10) is connected with ignoring the flip-flop transitions between different spins. From results of our recent paper [18] (see expressions (89) and (90) of that work), it follows that the characteristic time for flip-flop transitions is more than twice than the effective spin-spin relaxation time. This of course is prolonging the time interval where the described expressions can be reasonably accurate.

From our expressions it is clearly seen that the intermolecular magnetic dipole-dipole interactions contribute to the time dependence of the DQ NMR kinetic curves. Let us finally discuss in this respect the so-called normalized DQ intensity defined by Eq. (18) and having the approximate analytical expressions (22) and (23).

One can see that (23) is determined by the following function:

$$\tilde{G}(\tau_{DQ}) \equiv \frac{1}{N_s} \sum_{i,k} \langle \varphi_{ik}^{ex} \varphi_{ik}^{rec} \rangle, \qquad (24)$$

where the summation is performed over all spins in the system taking part in DQ NMR manipulations. Using expressions (2) and (9) and stationary property of lattice dynamical correlation functions, (24) can be rewritten in the following form:

$$\tilde{G}(\tau_{DQ}) \equiv \frac{\gamma^4 \hbar^2}{4} \int_0^{\tau_{DQ}} d\tau (\tau_{DQ} - \tau)(A_d(\tau_{DQ} + \tau) + A_d(\tau_{DQ} - \tau)), \qquad (25)$$

where

$$A_d(t) \equiv \frac{1}{N_s} \sum_{i,k} \left\langle \frac{(3\cos^2(\theta_{ik}(t)) - 1)}{r_{ik}^3(t)} \frac{(3\cos^2(\theta_{ik}(0)) - 1)}{r_{ik}^3(0)} \right\rangle. \qquad (26)$$

The correlation function (26) is nothing else than the total magnetic dipole-dipole correlation function. This function contains both intramolecular and intermolecular contributions. Since it has been proven, as was already mentioned in the Introduction, that the intermolecular part of the expression (26) plays an important role in the kinetic of the spin-lattice relaxation at frequencies of MHz and below, it is now clear from expressions (23-26) that it would also be important for DQ NMR phenomena. For polymer melts all components of $A_d(t)$ were analyzed in detail in our previous works [17,18] for different polymer dynamics models. Therefore, the expressions derived in these studies can without essential difficulties be exploited for the analysis of future experimental data connected with DQ NMR in polymer melts. The problem of taking into account intermolecular magnetic dipole-dipole interactions for DQ NMR kinetic curves is thus resolved in this paper in very general form with the accuracy of the second cumulant approximation.

Financial support from Deutsche Forschungsgemeinschaft (DFG) through grant STA 511/13-1 is gratefully acknowledged. The authors are grateful to R. Kimmich and K. Saalwächter for valuable discussions.